# Human Recognition based on Retinal Bifurcations and Modified Correlation Function

Amin Dehghani*
Department of Computer and
Electrical Engineering,
University of Tehran,
Tehran, Iran
dehghani@ut.ac.ir

* Corresponding author

*Abstract*— **Nowadays high security is an important issue for most of the secure places and recent advances increase the needs of high-security systems. Therefore, needs to high security for controlling and permitting the allowable people to enter the high secure places, increases and extends the use of conventional recognition methods. Therefore, a novel identification method using retinal images is proposed in this paper. For this purpose, new mathematical functions are applied on corners and bifurcations. To evaluate the proposed method we use 40 retinal images from the DRIVE database, 20 normal retinal image from STARE database and 140 normal retinal images from local collected database and the accuracy rate is 99.34 percent.**

## I. Introduction

Nowadays high security is an important issue for most of the secure places and recent advances increase the needs of high-security systems. Therefore, needs to high security for controlling and permitting the allowable people to enter the high secure places, increases and extends the use of conventional recognition methods.

There are different types of the biometric characteristics of a person. Some of well-known biometric characteristics for recognition as shown in Fig.1 are fingerprint, hand geometry, face, voice, signature, iris, and retina [1], [2]. Defrauding in recognition methods by using characteristics such as fingerprint or face is possible and it leads to the ineffectiveness of these methods for high secure places. Among all recognition methods, the only method that the users can't apply any modification for defrauding on biometric characteristic even after death is retinal recognition. Because retina is the innermost layer in the eye and two-dimensional images of visual objects are created on retina [3], [4] and there is not accessibility for people to change the pattern of retinal vessels which are the main biometric characteristic used for recognition. Depending on the application context, a biometric system may operate either in verification mode or identification mode. In the verification mode, the system validates a person's identity by comparing the captured biometric data with the own biometric template(s) stored system database. In such a system, an individual who desires to be recognized, claims an identity, usually via a PIN (Personal Identification Number), a user name, a smart card, etc., and the system conducts a one to one comparison to determine whether the claim is true or not. In the identification mode, the system recognizes an individual by searching the templates of all the users in the database for a match. Therefore, the system conducts a one-to-many comparison to establish an individual's identity (or fails if the subject is not enrolled in the system database) without the subject having to claim an identity [5]. The first identification system using commercial retina scanner called EyeDentification 7.5 was proposed by EyeDentify Company in 1976 [6], [7], [8] . In this paper, we propose a new method for human identification based on the corners and bifurcations extracted. Then, a modified correlation function is used to match the features of different people and finally for recognition. The rest of this paper is organized as follows. In section 2 we review last proposed methods for recognition based on the retinal images. Section 3 is devoted to Harris corner detector. Section 4 and 5 contains feature extraction, feature matching, and experimental results. Concluding remarks are given in the last section.

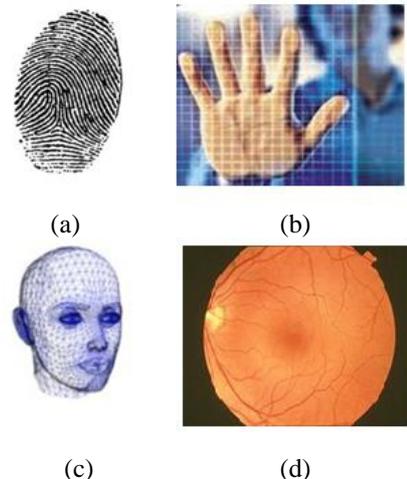

Figure 1. Biometric characteristics for identification: (a) finger print, (b) palm, (c) face, (d) retina [5].

## II. REVIEW OF LAST PROPOSED METHODS

Dehghani et al. [1], [8] used rotation invariant features and bifurcations for identification. For this purpose, some of the rotation invariant features of Hu's moments were extracted and after some steps such as transforming features and apply thresholding on the distance between features of different persons, the identity of testing persons were determined. Also using corners and bifurcation extracted Harris corner detector leaded to good identification rate.

Farzin et al. [9] proposed a novel method based on the features obtained from human retinal images. This system is composed of three principal modules including blood vessel segmentation, feature generation and feature matching. Blood vessel segmentation module has the role of extracting blood vessels pattern from retinal images. Feature generation module includes optical disc detection and selecting circular region around the optic disc of the segmented image. Then, using a polar transformation, a rotation invariant template is created. In the next stage, these templates are analyzed in three different scales using wavelet transform to separate vessels according to their diameter sizes. In the last stage, vessels position and orientation in each scale are used to define a feature vector for each subject in the database. For feature matching, they introduced a modified correlation measure to obtain a similarity index for each scale of the feature vector. Then, they compute the total value of the similarity index by summing scale-weighted similarity indices. Experimental results on a database, including 300 retinal images obtained from 60 subjects, demonstrated an average rate equal to 99 percent for the identification system.

Xu et al. [10] proposed a new method for recognition. They used the green grayscale ocular fundus image. The skeleton feature of optic fundus blood vessel using contrast-limited adaptive histogram equalization is extracted at first step. After filtering treatment and extracting shape feature, shape curve of blood vessels is obtained. Shape curve matching is later carried out by means of reference point matching. In their method for recognition, feature matching consists of finding affine transformation parameters which relate the query image and its best corresponding enrolled image. The computational cost of this algorithm is high because a number of rigid motion parameters should be computed for all possible correspondences between the query and enrolled images in the database. Experimental results on a database including 200 images resulted in zero false recognition against 38 false rejections.

Ortega et al. [11] used a fuzzy circular Hough transform to localize the optical disk in the retina images. Then, they defined feature vectors based on the ridge endings and bifurcations from vessels obtained from a crease model of the retinal vessels inside the optical disk. For matching, they used a similar approach as in [10] to compute the parameters of a rigid transformation between feature vectors which give the highest matching score. This algorithm is computationally more efficient with respect to the algorithm presented in [10]. This method used only for verification.

Tabatabaee et al. [12] presented a new algorithm based on the fuzzy C-means clustering algorithm. They used Haar wavelet and snakes model for optic disc localization. The Fourier-Mellin transform coefficients and simplified moments of the retinal image have been used as extracted features for the system. The computation cost and implementation time of this algorithm is high and the performance of the algorithm has been evaluated using a very small database including only 27 subjects.

Shahnazi et al. [13] proposed a new method based on the wavelet energy feature (WEF) which is a powerful tool of multi-resolution analysis. WEF can reflect the wavelet energy distribution of the vessels with different thickness and width in several directions at different wavelet decomposition levels (scales), so its ability to discriminate retinas is very strong. Easiness to compute is another virtue of WEF. Using semiconductors and various environmental temperatures in electronic imaging systems cause noisy images, so they used noisy retinal images for recognition. In existence of 5db to 20db noise, the proposed method can achieve 100% recognition rates on a database including 400 images

Oinonen et al. [14] proposed a novel method for verification based on minutiae features. The proposed method consists of three steps: blood vessel segmentation, feature extraction, and feature matching. In practice, vessel segmentation can be viewed as a preprocessing phase for feature extraction. After segmentation, the next step is to extract the vessel crossings together with their orientation information. These data are then matched with the corresponding ones from the comparison image. The method uses the vessel direction information for improved matching robustness.

## III. HARRIS CONER DETECTOR

In the Harris method, a window is slide on the image and the change of intensity resulted from the sliding window is determined [15]. Therefore, the change $E$ produced by displacement $(x,y)$ in the image intensity $I$ can be calculated using the following equation:

$$E(x,y) = \sum_u \sum_v w_{u,v}(I_{x+u,y+v} - I_{u,v})^2 \qquad (1)$$

Where $w$ is the moving window. Harris corner detector looks for local maxima in $\min\{E\}$ found over all considered moving directions [15]. Now, the amount of displacement is expressed using the following equation:

$$E(x,y) = \sum_u \sum_v w_{u,v}(I_{x+u,y+v} - I_{u,v})^2 = \sum_u \sum_v w_{u,v}(xX + yY + O(x^2,y^2))^2 \qquad (2)$$

where:

$$X = I * (-1,0,1) \approx \frac{\partial I}{\partial x}$$
$$Y = I * (-1,0,1)^T \approx \frac{\partial I}{\partial y} \qquad (3)$$

Other operators such as Sobel or Perwitt can be used for the gradient function in the above equation. Now for small displacement, E can be expressed as:

$$E(x,y) = Ax^2 + 2Cxy + By^2 \qquad (4)$$

where:

$$A = X^2 * w$$
$$B = Y^2 * w \qquad (5)$$
$$C = (XY) * w$$

To reduce the effect of the noise in the image, Gaussian window in the following form is used.

$$w(u,v) = e^{(u^2+v^2)/2\sigma^2} \qquad (6)$$

Finally, we can express $E$ as:

$$E(x,y) = (x,y)M(x,y)^T \qquad (7)$$

where $M$ is a symmetric matrix in following form:

$$M = \begin{bmatrix} A & C \\ C & B \end{bmatrix} \qquad (8)$$

$E$ is closely related to the local autocorrelation function. Let $\alpha$ and $\beta$ be the eigenvalues of $M$, therefore we have three situations. In Fig. 2, all regions such as edges, corners and regions with uniform intensity are described with respect to eigenvalues. In Figure 2, based on the $(\alpha, \beta)$ space, edges are in the regions in which one of the eigenvalues is very small and the other one is very large. Corners are in the regions where both eigenvalues are large and in flat regions both eigenvalues are small [15]. Harris and Stephens proposed the following corner measure:

$$R(x,y) = Det(M) - k(Tr(M))^2$$
$$Det(M) = \alpha \times \beta \qquad (9)$$
$$Tr(M) = \alpha + \beta$$

After extraction corners in the above equations, we use Localmax and thresholding (*Th*) method to extract final corners for identification. The parameters used in the above equations to extract corners and bifurcations are k=0.17 and $Th=7\times 10^4$. Therefore, using Harris corner detector, Localmax method and applying threshold, we extract corners and bifurcations for identification. To decrease the computational cost of our method, we use corners that their distance from the optic disc center is less than 80 pixels. For localizing the center of optic disc, we use template matching method proposed in [16]. Now we use the modified correlation proposed in [7] for identification.

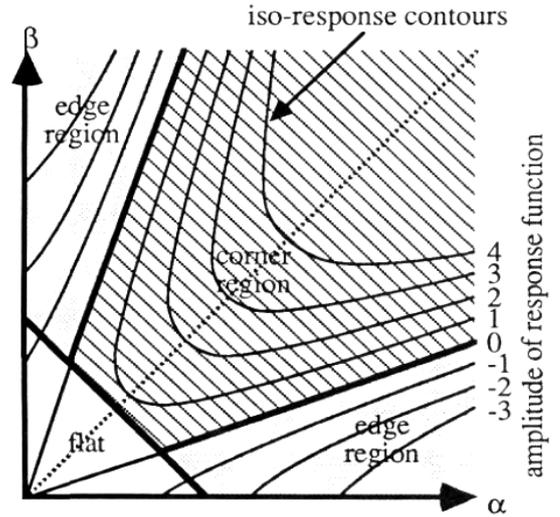

Figure 2. Auto-correlation principal curvature space, heavy lines give corner/edge/flat classification, fine lines are equi-response contours [15].

IV. FETURE VECTOR CONSTRUCTION

For feature construction, we use the similar method in [9], but in our proposed method to decrease the computational time, segmentation method is not used and instead of using vessels, we use corners and bifurcations. Based on the distance between corners and optic disc center in each retinal image, we classified the corners in three classes. The first class is allotted to corners that their distance from the center of optic disc is less than 25 pixels. The second class is for corner that their distances is more than 25 pixels and less than 50 pixels and the third class is for corners that their distances is more than 50 pixels and less than 80 pixels.

For constructing the feature vector, at first step we generate a pulse with the length of 360. Then, we localize corners in each class and replace them with rectangular pulses started at the orientation of the corner to optic disc center. The duration of each pulse for the first, second and third class is 2, 3 and 4 respectively, and its amplitude is equal to the orientation of the corner to the optic disc center.

The orientation of corners to optic disc center is between 0 and 360 degree. Each pulse for each corner is generated at the degree that the corner is located and it will repeat for all corners that their distance from optic disc center is less than 80 pixels. Therefore, the final feature vector is composed of 3 vectors (one per scale), each containing 360 values. Evidently, zero values in each vector correspond to none positions (orientations) of corners around the optic disc center. Therefore, for each retinal images we obtain $3 \times 360$ vectors.

V. FEATURE MATCHING

For feature matching, we use a similarity method proposed in [9]. The similarity between two feature vectors for the class is defined as follows:

$$Sim_i(\varphi) = \sum_{\tau=1}^{N} step(vec_i{}^{in}(\tau) \times$$
$$vec_i{}^{out}(\tau + \varphi)) \times \cos(2 \times (vec_i{}^{in}(\tau) \times \quad (10)$$
$$vec_i{}^{out}(\tau + \varphi)) \quad i = 1, 2, 3$$

Where $vec_i{}^{in}$ is the feature vector corresponding to the enrolled image, and $vec_i{}^{out}$ is the feature vector corresponding to the input query image. $N = 360$ is the length of the feature vector in each scale. step (·) is the step function defined as follows:

$$step(x) = \begin{cases} 1 & x > 0 \\ 0 & x \leq 0 \end{cases} \quad (11)$$

The role of step (·) in equation (11) is to normalize the product of pulse amplitudes in the feature vectors, because the amplitude of each pulse specifies the orientation of the corresponding vessel and is used only for determining the argument of cos (·) in equation (10). The role of cos (·) in (2) is to take into account the angle between vessels in the enrolled and query images. The similarity index between the enrolled and the query image corresponding to the $i$th class is defined as the maximum value of the modified correlation function:

$$SI_i = Max\{Sim_i(\varphi)\} \quad \varphi = \quad (12)$$
$$1, 2, \dots 360 \quad i = 1, 2, 3$$

Finally, a scale-weighted summation of SIs is computed to obtain a total SI for the enrolled and query images [6].

$$SI = w_1 \times SI_1 + w_2 \times SI_1 + w_3 \times SI_1 \quad (13)$$

Where SI is the total similarity index which is used for identification. In this work, To obtain the best weights, we did a global scanning of different weights and the best weights are: $w_1 = 1, w_2 = 2, w_3 = 4$.

*A. Expriments*

The proposed method is applied to a dataset containing 40 retina images from DRIVE database [17], 20 normal retinal image from STARE database and 140 normal retinal images from local collected database for testing our identification algorithm. For testing our algorithm, we organize 3 experiments. In the first experiment, we rotate the testing images (query images) 5 times, in the second experiment we rotate the testing images 10 times and in the final experiment we rotate the testing images 20 times. In Fig.3, we see some of retinal images, which are used in the proposed algorithm. In table 1, the results of experiments are shown.

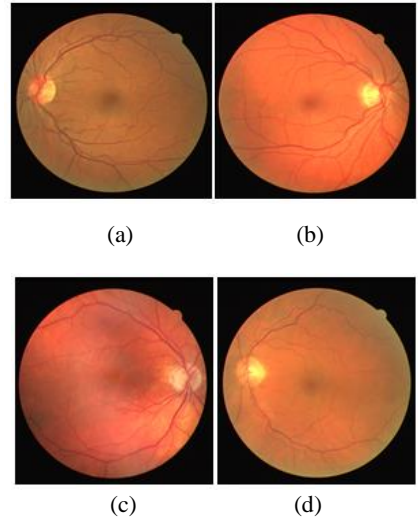

Figure 3: Some of the retinal images in dataset.

Table 1. Experimental results.

| Times of rotation | 5 | 10 | 20 | Mean |
|---|---|---|---|---|
| Accuracy | 99% | 99.2% | 99.5% | 99.34% |

The results of counterpart methods are shown in table 2.

Table 2. Comparing the results of different algorithms.

| Algorithms | Number of Images | Result |
|---|---|---|
| Farzin [9] | 300 | 99% |
| Xu [7] | - | 98.5% |
| Ortega [11] | 90 | 100% |
| Shahnazi et al. [13] | 400 | 100 |
| Oinonen [14] | 233 | 100% |
| Barkhoda et al. [18] | 360 | 98% |
| Proposed method | 4000 | 99.34% |

VI. CONCLUSION

In this paper, we proposed a new human identification method based on the corners and bifurcations and modified correlation function of retinal images. Therefore, Harris corner detector was used to obtain corners and bifurcations, then we used modified correlation function. The proposed algorithm has high accurate rate and thanks to avoiding preprocessing algorithms such as segmentation methods, it takes less computational time in comparison to the counterpart methods. The dataset used in this paper in the largest for retinal recognition and the accuracy of the proposed method for human identification is 99.34%.